\documentclass[prl,letterpaper,twocolumn,superscriptaddress,showpacs,preprintnumbers]{revtex4}

\usepackage{amsmath,amssymb,amsfonts,graphicx}

\RequirePackage{slashed}
\newcommand{\sh}[1]{\slashed{#1}}

\def\a{\alpha}
\def\g{\gamma}
\def\ve{\varepsilon}
\def\s{\sigma}
\def\o{\omega}

\def\D{\Delta}
\def\T{\Theta}

\def\pl{\partial}
\def\hs{\hspace}
\def\ol{\overline}
\def\no{\nonumber}

\def\lra{\longrightarrow}

\def\lf{\left}
\def\rg{\right}
\def\la{\langle}
\def\ra{\rangle}

\begin{document}

\preprint{NT@UW-09-03,~JLAB-THY-09-933}

\title{Isovector EMC effect explains the NuTeV anomaly}
\author{I.~C.~Clo\"et}
\email{icloet@phys.washington.edu}
\affiliation{Department of Physics, University of Washington, Seattle, WA 98195-1560, USA}
\author{W.~Bentz}
\affiliation{Department of Physics, School of Science, Tokai University,
             Hiratsuka-shi, Kanagawa 259-1292, Japan}
\author{A.~W.~Thomas}
\affiliation{Jefferson Lab, 12000 Jefferson Avenue, Newport News, VA 23606, USA and \\
College of William and Mary, Williamsburg, VA 23187, USA}

\begin{abstract}
A neutron or proton excess in nuclei
leads to an isovector-vector mean-field which, 
through its coupling to the quarks in a bound nucleon, implies a shift 
in the quark distributions with respect to the Bjorken scaling variable.
We show that this result leads to an additional correction to
the NuTeV measurement of $\sin^2\Theta_W$.
The sign of this correction is 
largely model independent and acts to reduce their result.
Explicit calculation within a covariant and confining Nambu--Jona-Lasinio model 
predicts that this vector field correction accounts for approximately
two-thirds of the NuTeV anomaly. We are therefore led to offer a new interpretation of
the NuTeV measurement, namely, that it is further evidence for the medium modification of the 
bound nucleon wavefunction.
\end{abstract}


\maketitle

Within relativistic, quark-level models of nuclear structure, the mean scalar and vector fields
in the medium generate fundamental changes in the internal structure of bound hadrons. 
These modifications lead to a good description of the EMC effect in finite nuclei and
predict a more dramatic modification of the bound nucleon spin
structure function \cite{Saito:1992rm,Cloet:2006bq,Cloet:2005rt}.
We show that in nuclei with $N \neq Z$ this approach leads to interesting and hitherto
unexplored effects connected with the isovector-vector mean-field,
which is usually represented by the $\rho^0$, and is in part responsible for the symmetry energy. 
In a nucleus like $^{56}$Fe or $^{208}$Pb where $N > Z$, the $\rho^0$ field will cause the $u$-quark 
to feel a small additional vector attraction and the $d$-quark to feel additional repulsion. 

In this Letter we explore the way in which this additional vector field
modifies the traditional EMC effect. However, there is an even more important
issue which is our main focus.
Even though the $\rho^0$ mean-field is completely consistent with charge symmetry,
the familiar assumption that $u_p(x) = d_n(x)$ and $d_p(x) = u_n(x)$ will clearly fail for a
nucleon bound in a nucleus with $N \neq Z$.
Therefore correcting for the $\rho^0$ field 
is absolutely critical in a situation where symmetry arguments are essential, 
such as the use of $N \neq Z$ nuclear data from $\nu$ and $\ol{\nu}$ deep inelastic 
scattering (DIS) to extract $\sin^2\Theta_W$ via the Paschos-Wolfenstein relation  \cite{Paschos:1972kj}. 
Indeed, we show that the deviation from the naive application of charge symmetry
to the $\nu$ and $\ol{\nu}$ data on $^{56}$Fe naturally explains the famous
NuTeV anomaly. 

The Paschos-Wolfenstein (PW) ratio is defined by \cite{paschos}
\begin{align}
R_{PW} = \frac{\s_{NC}^{\nu\,A} - \s_{NC}^{\bar{\nu}\,A}}{\s_{CC}^{\nu\,A} - \s_{CC}^{\bar{\nu}\,A}},
\label{eq:PW}
\end{align}
where $A$ represents the target, $NC$ indicates weak neutral current
and $CC$ weak charged current interaction. Expressing the cross-sections in terms of quark 
distributions and ignoring heavy flavour contributions, the PW ratio becomes
\begin{align}
\hs{-1.2mm}R_{PW} = \tfrac{\lf(\tfrac{1}{6}-\tfrac{4}{9} \sin^2\Theta_W\rg) \lf\la x_A\,u^-_A\rg\ra 
        + \lf(\tfrac{1}{6}-\tfrac{2}{9} \sin^2\Theta_W\rg)\, \lf\la x_A\,d^-_A \rg\ra}
       {\lf\la x_A\,d^-_A \rg\ra - \tfrac{1}{3}\lf\la x_A\,u^-_A \rg\ra},
\label{eq:PW_quark}
\end{align}
where $x_A$ is the Bjorken scaling variable of the nucleus multiplied by $A$,
$\la\ldots\ra$ implies integration over $x_A$,
and $q_A^- \equiv q_A - \bar{q}_A$ are the non-singlet quark distributions of the target.
Therefore, the quantities in the angle brackets are simply the momentum fractions of 
the target carried by the valence quarks. 
 
Ignoring quark mass differences and possible electroweak corrections the $u$- and 
$d$-quark distributions of an isoscalar target will be identical, and in this limit
Eq.~\eqref{eq:PW_quark} becomes
\begin{align}
R_{PW} \stackrel{N=Z}{\lra} \frac{1}{2} - \sin^2\Theta_W.
\label{eq:isoscalarPW}
\end{align}
If corrections to Eq.~\eqref{eq:isoscalarPW} are small the PW ratio provides a unique
way to measure the Weinberg angle.

Motivated by Eq.~\eqref{eq:isoscalarPW} the NuTeV collaboration extracted a value of $\sin^2\Theta_W$
from neutrino and anti-neutrino DIS on an iron target \cite{Zeller:2001hh}, finding
\begin{align}
\sin^2\Theta_W = 0.2277 \pm 0.0013(\text{stat.}) \pm 0.0009(\text{syst.}).
\end{align}
The three-sigma discrepancy between this result and the world average \cite{Abbaneo:2001ix},
namely $\sin^2\Theta_W = 0.2227 \pm 0.0004$, is the NuTeV anomaly. 
Some authors have speculated that the NuTeV anomaly supports the existence of physics
beyond the Standard Model \cite{Davidson:2001ji}. However, existing high precision data for 
other electroweak observables places tight constraints on new physics explanations.
Standard Model corrections to the NuTeV result have largely been focused on
nucleon charge symmetry violating effects~\cite{Sather:1991je} 
and a non-perturbative strange quark sea \cite{Davidson:2001ji}. Charge symmetry violation, arising from the 
$u$- and $d$-quark mass
differences, is probably the best understood and constrained correction and 
can explain approximately one-third of the NuTeV anomaly~\cite{Londergan:2003pq}. 
Standard nuclear corrections like Fermi motion and binding are found to be small \cite{Kulagin:2003wz}. 
However effects from the medium modification of the bound nucleon, which are now 
widely accepted as an essential ingredient in explaining 
the EMC effect~\cite{Smith:2002ci}, 
have hitherto not been explored in relation to the NuTeV anomaly.

In our approach, presented in Refs.~\cite{Mineo:2003vc,Cloet:2005rt,Cloet:2006bq}, the
scalar and vector mean-fields inside a nucleus couple to the quarks in the bound nucleons and 
self-consistently modify their internal structure. The scalar field
renormalizes the constituent quark mass, resulting in effective hadron masses in-medium.
The influence of the vector fields on the quark distributions arises from the non-local nature 
of the quark bilinear in their definition~\cite{Mineo:2003vc}. This leads to  
a largely model independent result for the modification of the in-medium parton distributions 
of a bound nucleon by the vector mean-fields \cite{Steffens:1998rw,Mineo:2003vc,Detmold:2005cb}, namely
\begin{align}
q(x) = \frac{p^+}{p^+-V^+}\ q_{0}\lf(\frac{p^+}{p^+-V^+}\,x - \frac{V_q^+}{p^+-V^+}\rg) .
\label{eq:shift}
\end{align}
The subscript 0 indicates the absence of vector fields and $p^+$ is the nucleon lightcone plus 
component of momentum. The quantities $V^+$ 
and $V_q^+$ are the lightcone plus component of the net vector field felt by the nucleon
and a quark of flavour $q$, respectively. 
Importantly Eq.~\eqref{eq:shift} is consistent with baryon number and momentum conservation,
and implies that the mean vector field carries no momentum.

Before embarking on explicit calculations, we first explore the model 
independent consequences of Eq.~\eqref{eq:shift} for the
PW ratio and the subsequent NuTeV measurement  of $\sin^2\T_W$.
The NuTeV experiment was performed on a predominately $^{56}$Fe target, 
and therefore isoscalarity corrections need to be applied to the PW ratio 
before extracting $\sin^2\T_W$. Isoscalarity corrections to Eq.~\eqref{eq:isoscalarPW} 
for small isospin asymmetry have the general form
\begin{align}
\D R_{PW} \simeq \lf(1-\frac{7}{3}\sin^2\Theta_W\rg)
                        \frac{\la x_A\,u^-_A - x_A\,d^-_A\ra}{\la x_A\,u^-_A + x_A\,d^-_A\ra},
\label{eq:iso_correct}
\end{align}
where the $Q^2$ dependence of this correction resides completely with $\sin^2\Theta_W$.
NuTeV perform what we term naive isoscalarity corrections, where 
the neutron excess correction is determined by assuming the target is 
composed of free nucleons \cite{nuteviso}.
However there are also isoscalarity corrections from medium effects, in particular from
the medium modification of the structure functions of \textit{every} nucleon in the 
nucleus, arising from  the isovector $\rho^0$ field.
For nuclei with $N > Z$ the $\rho^0$ field develops a non-zero expectation value that results
in $V_u < V_d$, that is, the $u$-quarks feel less vector repulsion than the $d$-quarks.
A direct consequence of this and the transformation given in Eq.~\eqref{eq:shift} is that
there must be a small shift in quark momentum from the $u$- to the $d$-quarks.
Therefore the momentum fraction $\la x_A\,u^-_A - x_A\,d^-_A\ra$ in Eq.~\eqref{eq:iso_correct} will be 
negative, even after naive isoscalarity corrections are applied. Correcting for the 
$\rho^0$ field will therefore have the \textit{model independent effect of reducing the 
NuTeV result for $\sin^2\Theta_W$}.
As we shall see, this correction largely explains the NuTeV anomaly.

To determine the nuclear quark distributions we
use the Nambu--Jona-Lasinio (NJL) 
model \cite{Nambu:1961tp}, which is viewed as a low energy chiral effective 
theory of QCD and is characterized by a 4-fermion contact interaction between the quarks.
The NJL model has a long history of success in describing mesons as $\bar{q}q$ bound states \cite{Vogl:1991qt}
and more recently as a self-consistent model for free and in-medium baryons 
\cite{Cloet:2006bq,Cloet:2005rt,Cloet:2005pp,Mineo:2003vc}. 
The original 4-fermion interaction term in the NJL Lagrangian can be decomposed into various 
$\bar{q}q$  and $qq$ interaction channels via Fierz transformations \cite{Ishii:1995bu}. 
The relevant terms of the NJL Lagrangian to this discussion are
\begin{align}
\mathcal{L} &= \ol{\psi}\lf(i\sh{\pl} - m\rg)\psi
+\,G_\pi\lf(\lf(\ol{\psi}\psi\rg)^2 - \lf(\ol{\psi}\g_5\,\vec{\tau}\psi\rg)^2\rg)   \no \\
&\hs{1mm}-\,G_\o\lf(\overline{\psi}\g^\mu\psi\rg)^2 -\,G_\rho\lf(\ol{\psi}\g^\mu\vec{\tau}\psi\rg)^2 \no \\
&\hs{1mm}+\,G_s \Bigl(\ol{\psi}\,\g_5 C \tau_2 \beta^A\, \ol{\psi}^T\Bigr)
                               \Bigl(\psi^T\,C^{-1}\g_5 \tau_2 \beta^A\, \psi\Bigr) \no \\
&\hs{1mm}+\,G_a \lf(\ol{\psi}\,\g_\mu C \tau_i\tau_2 \beta^A\, \ol{\psi}^T\rg) 
                               \Bigl(\psi^T\,C^{-1}\g^{\mu} \tau_2\tau_i \beta^A\, \psi\Bigr),
\label{eq:lag}
\end{align}
where $\beta^A = \sqrt{\tfrac{3}{2}}\,\lambda^A~(A=2,5,7)$ are the 
the colour $\ol{3}$ matrices~\cite{Cloet:2006bq}, $C = i\g_2\g_0$ and 
$m$ is the current quark mass.

The scalar $\bar{q}q$ interaction term generates the scalar field, which dynamically
generates a constituent quark mass via the gap equation.
The vector $\bar{q}q$ interaction terms
are used to generate the isoscalar-vector, $\omega_0$, and isovector-vector, $\rho_0$, mean-fields 
in-medium. The $qq$ interaction terms give the diquark
$t$-matrices whose poles correspond to the scalar and axial-vector diquark masses. 
The nucleon vertex function and mass are obtained by solving the homogeneous 
Faddeev equation for a quark and a diquark, where the static approximation is used to truncate the
quark exchange kernel \cite{Cloet:2005pp}. To regularize the NJL model we choose the
proper-time scheme, which enables the removal of unphysical thresholds for nucleon decay 
into quarks, and hence simulates an important aspect of confinement \cite{Ebert:1996vx,Bentz:2001vc}.

To self-consistently determine the strength of the mean scalar and vector fields,
an equation of state for nuclear matter is derived from the NJL Lagrangian, Eq.~\eqref{eq:lag},
using hadronization techniques~\cite{Bentz:2001vc}. In a mean-field approximation
the result for the energy density is~\cite{Bentz:2001vc}
\begin{equation}
\mathcal{E} = \mathcal{E}_V - \frac{\o_0^2}{4G_\o} - \frac{\rho_0^2}{4G_\rho} + \mathcal{E}_p + \mathcal{E}_n,
\label{eq:effective}
\end{equation}
where the vacuum energy $\mathcal{E}_V$ has the familiar Mexican hat shape and the
energies of the protons and neutrons  moving through the mean scalar and vector fields are labelled
by $\mathcal{E}_p$ and $\mathcal{E}_n$, respectively. The corresponding proton and neutron Fermi energies are
\begin{align}
\ve_{F\a} \!= E_{F\a}\!\! + V_{\a}\! =\! \sqrt{M_N^{*2} + p_{F\a}^2} + 3\o_0 \pm \rho_0,
\end{align}
where $\a = p$ or $n$, the plus sign refers to the proton, $M_N^*$ is the in-medium nucleon 
mass and $p_{F\a}$ the nucleon Fermi momentum.
Minimizing the effective potential 
with respect to each vector field gives the following useful relations:
$\omega_0 = 6\,G_\omega\lf(\rho_p + \rho_n\rg)$ and $\rho_0   = 2\,G_\rho\lf(\rho_p - \rho_n\rg)$,
where $\rho_p$ is the proton and $\rho_n$ the neutron density.
The vector field experienced by each quark flavour is given by
$V_u = \omega_0 + \rho_0$ and $V_d = \omega_0 - \rho_0$.

\begin{figure}[tbp]
\centering\includegraphics[width=\columnwidth,clip=true,angle=0]{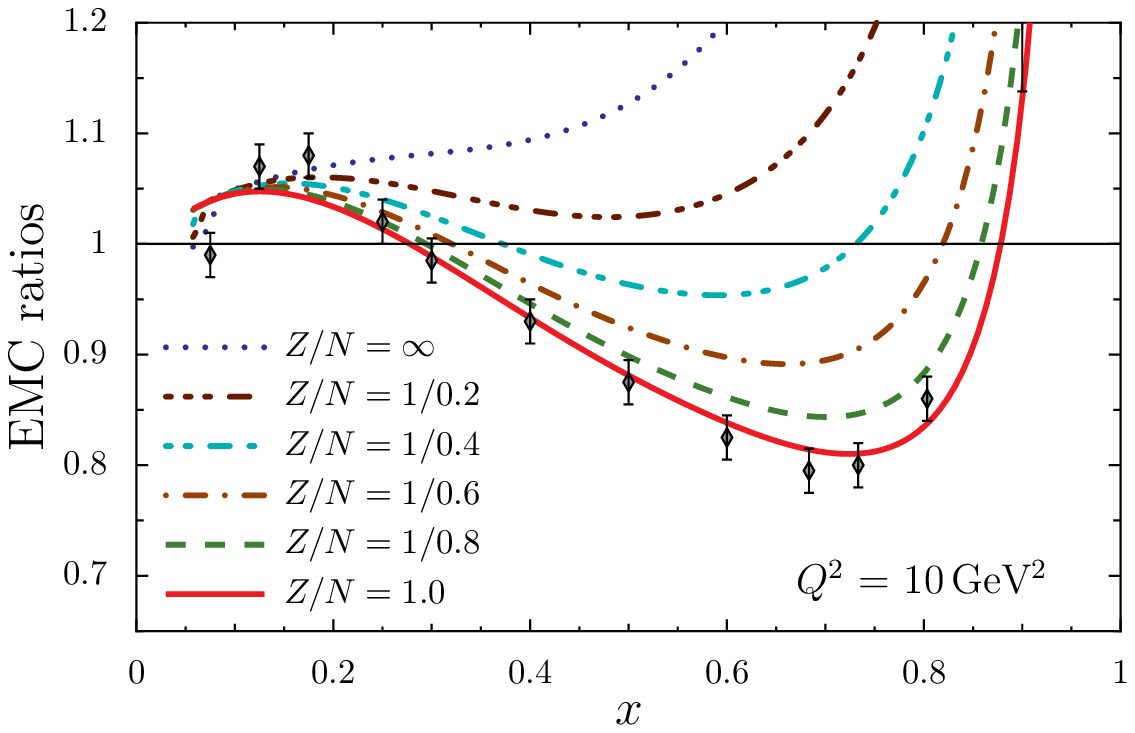}
\caption{Isospin dependence of the EMC effect for proton-neutron ratios greater than one.
The data is from Ref.~\cite{Sick:1992pw} and corresponds to $N=Z$ nuclear matter.}
\label{fig:ratio1}
\vspace{-1.0em}
\end{figure}

As explained in Ref.~\cite{Cloet:2006bq}, the parameters of the model 
are determined by standard hadronic properties, and the empirical saturation energy 
and density of symmetric nuclear matter. The new feature of this work is the
$\rho^0$ field, where $G_\rho$ is determined by the empirical symmetry energy of 
nuclear matter, namely $a_4=32\,$MeV, giving $G_\rho = 14.2\,$GeV$^{-2}$. 

Details of our results for the free and $N\simeq Z$ in-medium parton distributions are given in 
Refs.~\cite{Cloet:2005pp,Cloet:2006bq,Cloet:2005rt}. 
For in-medium isospin dependent parton distributions our produce is
as follows: Effects from the scalar mean-field are included by replacing
the free masses with the effective masses in the expressions for the free parton distributions 
discussed in Ref.~\cite{Cloet:2005pp}. For asymmetric nuclear matter ($N\neq Z$) the proton
and neutron Fermi momentum will differ and therefore so will their Fermi smearing functions.
To include the nucleon Fermi motion, the quark distributions modified by the
scalar field are convoluted with the appropriate Fermi smearing function, namely
\begin{align}
\hs{-1.5mm}f_{\a0}(y_A) = \tfrac{\mathcal{N}\a}{A}\, \tfrac{3}{4} \lf(\tfrac{\hat{M}_N}{p_{F\a}}\rg)^{\!3}\! 
             \lf[\lf(\tfrac{p_{F\a}}{\hat{M}_N}\rg)^{\!2}\! - \lf(\tfrac{E_{F\a}}{\hat{M}_N}-y_A\rg)^{\!2}\rg]\!,
\label{eq:smearing}
\end{align}
where $\mathcal{N}_p = Z$, $\mathcal{N}_n = N$ and $\hat{M}_N = \tfrac{Z}{A}\,E_{Fp} + \tfrac{N}{A}\,E_{Fn}$. 
Vector field effects can be included in Eq.~\eqref{eq:smearing} by
the substitutions $E_{F\a} \to \ve_{F\a}$ and 
$\hat{M}_N \to \ol{M}_N = \tfrac{Z}{A}\,\varepsilon_{Fp} + \tfrac{N}{A}\varepsilon_{Fn}$.
Our final result for the infinite asymmetric nuclear matter quark distributions,
which includes vector field effects on both the quark distributions in the bound nucleon 
and on the nucleon smearing functions, is given by
\begin{align}
q_{A}(x_A) &= \frac{\ol{M}_N}{\hat{M}_N}\ q_{A0} \lf(\frac{\ol{M}_N}{\hat{M}_N} x_A - \frac{V_{q}}{\hat{M}_N}\rg).
\label{eq:shift2}
\end{align}
The subscript $A0$ indicates a distribution which includes effects from Fermi motion and the 
scalar mean-field. 
The distributions calculated in this way are then evolved \cite{Miyama:1995bd} from the model 
scale, $Q_0^2 = 0.16\,$GeV$^2$, to an appropriate $Q^2$ for comparison with experimental data.

\begin{figure}[tbp]
\centering\includegraphics[width=\columnwidth,clip=true,angle=0]{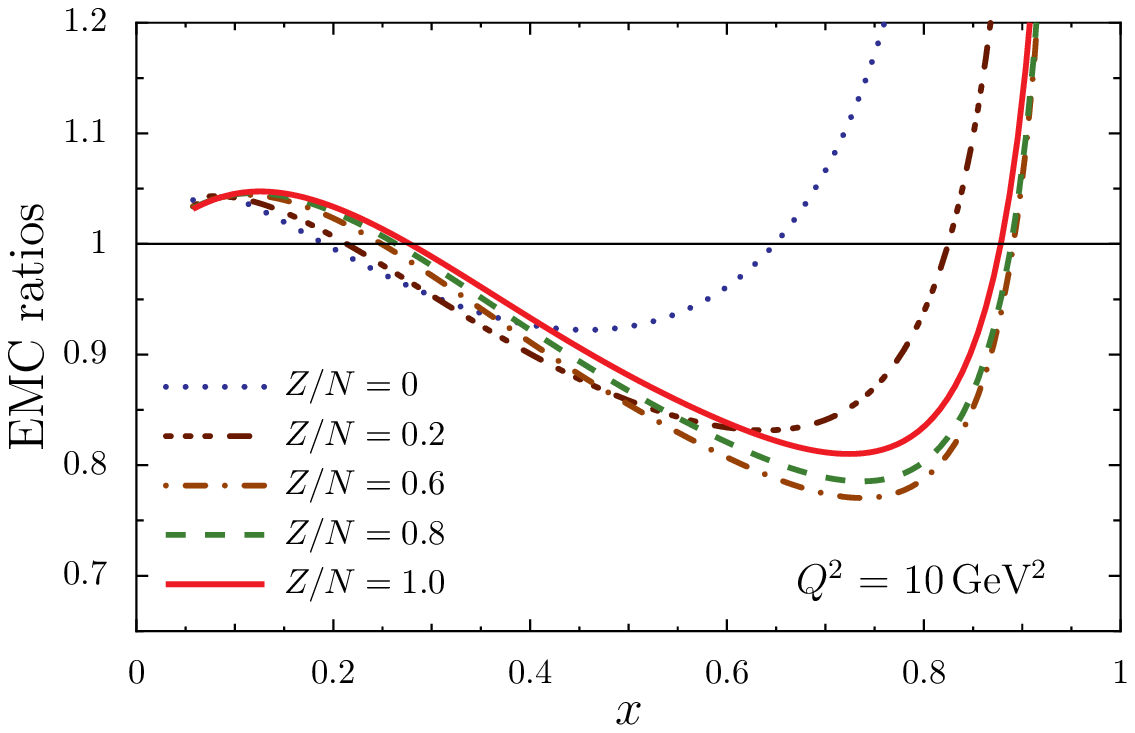}
\caption{Isospin dependence of the EMC effect for proton-neutron ratios less than one.}
\label{fig:ratio2}
\vspace{-1.0em}
\end{figure}

The EMC effect is defined by the ratio
\begin{align}
R = \frac{F_{2A}}{F_{2A}^{\text{naive}}} = \frac{F_{2A}}{Z\,F_{2p} + N\,F_{2n}} 
                                      \simeq \frac{4u_A + d_A}{4u_f + d_f},
\end{align}
where $q_A$ are the quark distributions of the target and $q_f$ are the distributions of the 
target if it was composed of free nucleons. Results for the isospin dependence of the EMC 
effect are given in Figs.~\ref{fig:ratio1} and \ref{fig:ratio2}. 

Fig.~\ref{fig:ratio1} illustrates the EMC effect for proton rich matter, where we find a
decreasing effect as $Z/N$ increases. An intuitive understanding of this result
may be obtained by realizing that it is a consequence of binding effects at the quark level. 
For $Z/N > 1$ the $\rho_0$ field is positive, which means $V_u > V_d$ and hence 
the $u$-quarks are less bound than the $d$-quarks.
Therefore the $u$-quark distribution becomes less modified while medium modification of the 
$d$-quark distribution is enhanced. Since the EMC effect is dominated by the 
$u$-quarks it decreases. The isospin dependence of the  EMC effect for nuclear matter 
with $Z/N < 1$ is given in Fig.~\ref{fig:ratio2}. Here the medium modification
of the $u$-quark distribution is enhanced, while the $d$-quark distribution is modified
less by the medium. Since the EMC ratio is initially dominated by the $u$-quarks
the EMC effect first increases as $Z/N$ decreases from one. However, eventually the 
$d$-quark distribution dominates the ratio and at this stage the EMC effect begins to
decrease in the valence quark region. We find a maximal EMC effect for $Z/N \simeq 0.6$, 
which is slightly less than the proton-neutron ratio in lead. This isospin dependence is clearly 
an important factor in understanding the $A$ dependence of the EMC effect, even after standard
neutron excess corrections are applied.

Now we turn to the consequences of the isospin dependence of the EMC effect
for the NuTeV measurement of $\sin^2\Theta_W$.
The NuTeV experiment was performed on an iron target, which, because of impurities
had a neutron excess of 5.74\% \cite{Zeller:2001hh}. Choosing our $Z/N$ ratio to
give the same neutron excess, we use our medium modified quark distributions
and Eq.~\eqref{eq:iso_correct} to determine the full isoscalarity correction to
the isoscalar PW ratio, given by Eq.~\eqref{eq:isoscalarPW}. Using the Standard Model value 
for the Weinberg angle we obtain $\D R_{PW} = -0.0139$.
If we break this result into the three separate isoscalarity 
corrections, by using Eq.~\eqref{eq:iso_correct} and the various stages
of the medium modified quark distributions, we find 
\begin{align}
\D R_{PW} &= \D R^{\text{naive}}_{PW} + \D R^{\text{Fermi}}_{PW} + \D R^{\raisebox{1.0pt}{${\scriptstyle\rho}$}^0}_{PW} \no \\
         &= -\lf(0.0107 + 0.0004 + 0.0028\rg).
\end{align}
Higher order corrections to Eq.~\eqref{eq:iso_correct} do not change this result.
The NuTeV analysis includes the naive isoscalarity correction~\cite{nuteviso} but is missing the medium correction
of $-0.0032$ \cite{functional}. This new correction accounts for two-thirds of the NuTeV anomaly. If
we also include the well constrained charge symmetry violation (CSV) correction,
$\D R_{PW}^{CSV} = -0.0017$~\cite{Londergan:2003pq}, which originates from 
the quark mass differences, we have a total correction of 
$\D R_{PW}^{\text{medium}} + \D R_{PW}^{CSV} = -0.0049$.
The combined correction accounts for the NuTeV anomaly \cite{nutevcorrect}.

Since CSV and medium modification corrections explain 
the discrepancy between the NuTeV result and the Standard Model,
we propose that this NuTeV measurement provides strong evidence that the nucleon
is modified by the nuclear medium, and should not be interpreted as an indication 
of physics beyond the Standard Model. In our opinion this conclusion is equally
profound since it may have fundamental consequences for our understanding of traditional 
nuclear physics. We stress that the physics presented in this 
paper, in particular the effects of the $\rho^0$ mean-field, are consistent with existing 
data~\cite{Kumano:2002ra}, 
but can strongly influence other observables. For example, the $\rho^0$ field gives rise to a 
strong flavour dependence of the EMC effect, and experimental proposals have been submitted 
at Jefferson Lab to look for such effects.

I.~C. thanks Jerry Miller for helpful discussions.
This work was supported by the the U.S. Department of Energy under Grant No. DEFG03-97ER4014
and by Contract No. DE-AC05-06OR23177, under which Jefferson Science Associates,
LLC operates Jefferson Laboratory and by the Grant in Aid for Scientific
Research of the Japanese Ministry of Education, Culture, Sports, Science and
Technology, Project No. C-19540306.


\begin{thebibliography}{40}
\bibitem{paschos}
  The cross-sections in Eq.~\eqref{eq:PW} have been integrated over the Bjorken scaling
  variable and energy transfer. The $Q^2$ dependence of the the PW ratio resides
  with $\sin^2\Theta_W$.
\bibitem{nuteviso}
  NuTeV do not directly utilize Eq.~\eqref{eq:iso_correct} for their naive isoscalarity correction, 
  because in their case, details of this correction depend explicitly on the Monte-Carlo routine used 
  to analyze their data.
\bibitem{nutevcorrect}
  The NuTeV result for $R_{PW}$ after naive isoscalarity correction is $0.2723$
  which differs from $\tfrac{1}{2}-\sin^2\T_W =0.2773$ by the amount $-0.005$.
\end{thebibliography}
\end{document}